\documentclass[twocolumn,aps,prb,graphicx,showpacs]{revtex4}

\usepackage{graphicx}

\begin{document}

\title{Solution of the local field equations for self-generated glasses}
\author{Sangwook Wu$^{(a)}$, J\"org Schmalian$^{(a)}$, Gabriel Kotliar$%
^{(b)} $, and Peter G Wolynes$^{(c)}$.}
\affiliation{$^{(a)}$Department of Physics and Astronomy and Ames Laboratory, Iowa State
University, Ames, IA 50011\\
$^{(b)}$ Serin Physics Laboratory, Rutgers University, 136 Freylinghuysen
Road, Piscataway, NY 08854\\
$^{(c)}$ Department of Chemistry and Biochemistry, University of California,
Dan Diego, La Jolla, CA 92093 }
\date{\today}

\begin{abstract}
We present a self-consistent local approach to self generated glassiness
which is based on the concept of the dynamical mean field theory to many
body systems. Using a replica approach to self generated glassiness, we map
the problem onto an effective local problem which can be solved exactly.
Applying the approach to the Brazovskii-model, relevant to a large class of
systems with frustrated micro-phase separation, we are able to solve the
self-consistent local theory without using additional approximations. We
demonstrate that a glassy state found earlier in this model is generic and
does not arise from the use of  perturbative approximations. In addition we
demonstrate that the glassy state depends strongly on the strength of the
frustrated phase separation in that model.
\end{abstract}

\maketitle

\section{Introduction}

The familiar example of vitrification upon super-cooling molecular fluids
provides proof that an effectively non-ergodic disordered state can be
generated in a system without pre-existing quenched disorder. Such
self-generated glassiness may be a more widespread possibility in the
material world than is currently acknowledged. \ Candidate solid state
systems with extremely slow dynamics often have impurities remaining from
their synthesis so there is always a suspicion that non-ergodicity comes from
disorder.\cite%
{mang01,Millis96,Dagotto,nmr01,nmr02,nmr03,nmr3b,nmr04,msr01,msr02,msr03}
While a rather successful microscopic theory of molecular glasses now exists
based on the idea of an underlying random first order transition\cite{KTW89a,KTW89b,KTW89c,KTW89d}, this theory, like all theories of liquid state phenomena, contains
approximations that can only be checked {\em a  posteriori } through their agreement
with simulations or experiments. To deepen and widen our understanding of
glassy dynamics beyond the molecular fluid example, it would be desirable to
have a model which can be analyzed in a formally exact fashion
using familiar tools of continuum field theory. Such a theory would also
help clarifying which features of the current theory of structural glasses
are robust and "protected"\cite{TOE} and which are fortuitously correct.

Both, the recent treatment of the thermodynamics of fragile glasses
developed within a replica formulation\cite{MP991} and the closely related
approach using  density functional theory for super-cooled liquids\cite{SSW85} \
suggest that the major aspects of glass formation  \ stem from the very
strong local correlations of a dense super-cooled fluid. This is very
different \ from systems with critical dynamics close to second order phase
transitions, where long wavelength fluctuations are
paramount. One finds experimentally that dynamics on essentially all
wavelengths slows to nearly the same extent as the glass transition is
approached. This suggests that the search for an solvable limit of the glass
problem should focus on the strong but local nonlinearities. 

Major progress has recently been made in the theoretical understanding of
strongly interacting systems with predominantly local correlations. Strongly
correlated electron systems like Mott insulators or systems with strong
electron-phonon interactions have been investigated with the dynamic mean
field theory\cite{dmft1,dmft2} (DMFT). Originally, this approach was based
on the assumption that all non-trivial correlations of a given system are
strictly local (for the precise definition of a local  theory, see below)
and then allowed for a solution of such systems without further
approximations. More recently several generalizations of the approach have
been developed, which allow the inclusion of collective excitations\cite%
{Si,CK90} as well as short range correlations in so called cluster DMFT
theories\cite{cluster2,cluster3,cluster}.

In this paper we will follow the main strategy of the dynamic mean field
approach and apply it to the replica theory of self generated glassiness.
This allows us to solve completely for the long time correlation function
(Edwards-Anderson parameter) as well as the configurational entropy of a
glass within a local theory. Even though the origin of glassiness is not
quenched disorder, there is a similarity between our theory and the mean
field theory of spin glasses which is also exact in the limit of a purely
local approach. Solvable local limits like the  Sherrington-Kirkpatrick model,
turned out to be extremely fruitful for the investigation of more realistic
models. To identify and solve such a local problem is therefore of general
interest. Another appealing aspect of the approach developed here is that
the recent generalizations of the DMFT to cluster systems offer a feasible
way to  develop a controlled series of approximations for candidate glassy
systems that successively extend the range of local correlations taken into
account.\cite{cluster2,cluster3,cluster}. Clearly, before this becomes possible we have to set up and analyze
the strictly local (i.e. mean field) theory. The solutions of the local
field theories discussed in this paper rely on a self consistency condition
that is uniformly applied to the replicated statistical mechanics problem.
The droplet effects that give rise to the activated transitions in molecular
fluids cannot be described by this uniform replica symmetry breaking ansatz.
While the self consistent replica field theory is applicable to account for
some geometric effects in finite dimensions the latter highly
nonperturbative effects are absent, but are suppressed if the coordination
number is high. We should also mention  that related interesting approach to
self-generated glassiness in a model in infinite dimensions (where the
locality of the problem is exactly guaranteed) that was recently presented by
Lopatin and Ioffe.\cite{Lopatin} 

We demonstrate that a uniformly frustrated system with competition of
interactions on different length scales undergoes  a glass transition once
inhomogeneous fluctuations become strong enough if no ordered phase forms via nucleation.
 Specific calculations are carried out  for the case of the Brazovskii model\cite{Brazovskii}, which
describes systems with tendency towards micro-phase separation. Glassiness
was shown to exist in this model in Refs.\onlinecite{SW00,SWW00}. However, it
remained unclear to what extent this is merely a consequence of an
approximate solution using perturbation theory or is indeed the correct mean
field behavior of this model.

In the next section we briefly summarize the replica approach to self
generated glassiness, present the main idea of the dynamic mean field theory
and solve the local theory for self generated glassiness in case of the
Brazovskii model. We \ summarize the results of this paper in Section III.

\section{Theory}

\subsection{The replica approach}

In this paper we consider physical systems consisting of $\ L$-components
(types of atoms or molecules) and with short ranged $3^{\mathrm{rd}}$, $4^{%
\mathrm{th}}$ and higher \ order virial coefficients. No specific assumption
is made with respect to the (possibly non local) \ $2^{\mathrm{nd}}$ virial
coefficient. \ In terms of a density field $\rho _{l}\left( \mathbf{x}%
\right) =\sum_{i}\delta \left( \mathbf{x-X}_{i,l}\right) $ of particles of
type $l=1,...,L$ ( $L=1$ in case of a one component system) we then have an
effective energy: 
\begin{equation}
H\left[ \phi \right] =H_{0}\left[ \phi \right] +\int d^{d}xV\left( \phi
_{l}\left( \mathbf{x}\right) \right)   \label{hamgen}
\end{equation}

\bigskip 

\bigskip 

with 
\begin{equation}
H_{0}=\int d^{d}xd^{d}x^{\prime }\sum_{ll^{\prime }}\phi _{l}\left( \mathbf{x%
}\right) G_{0ll^{\prime }}^{-1}\left( \mathbf{x},\mathbf{x}^{\prime }\right)
\phi _{l^{\prime }}\left( \mathbf{x}^{\prime }\right) .  \label{hamgen2}
\end{equation}%
where $\phi _{l}\left( \mathbf{x}\right) =\rho _{l}\left( \mathbf{x}\right) -%
\overline{\rho }_{l}$ is the deviation of the density from its average
value. The key assumption made in Eqs.\ref{hamgen} and \ref{hamgen2} is the
local character of $V\left( \phi \right) $ which includes the higher order
virial coefficients. For example, Eq.\ref{hamgen}, might be a representation
of the density functional used by Ramakrishnan and Youssoff with $%
G_{0ll^{\prime }}^{-1}\left( \mathbf{x},\mathbf{x}^{\prime }\right)
=c_{ll^{\prime }}\left( \mathbf{x}-\mathbf{x}^{\prime }\right) $ the direct
correlation function of the fluid\cite{Hansen} and 
\begin{equation}
V\left( \phi \right) =\sum_{l}\left( \overline{\rho }_{l}+\phi _{l}\left( 
\mathbf{x}\right) \right) \left[ \log \left( \overline{\rho }_{l}+\phi
_{l}\left( \mathbf{x}\right) \right) -1\right] 
\end{equation}%
the ideal gas free energy.

The partition function of the system is given by 
\begin{equation}
Z_{\mathrm{eq}}=\int D\phi \exp \left( -H\left[ \phi \right] /T\right)
\end{equation}
and determines the equilibrium properties of the system. In a system with
complex energy landscape where we expect that the system might undergo
vitrification, the knowledge of the equilibrium partition function is not
sufficient anymore.

A widely accepted view\cite{KTW89a,KTW89b} is that a glassy system may be \
considered to be  trapped in local metastable states for very long time and
can therefore not realize a considerable part of the entropy of the system,
\ called the configurational entropy $S_{\mathrm{c}}=\log \mathcal{N}_{%
\mathrm{ms}}$, where $\mathcal{N}_{\mathrm{ms}}$ is the number of metastable
states. If $\mathcal{N}_{\mathrm{ms}}$ is exponentially large with respect
to the size of the system, $S_{\mathrm{c}}$ becomes extensive and ordinary
equilibrium thermodynamics fails. There are several theoretical approaches
which offer a solution to this breakdown of equilibrium many body theory. On
the one hand one can solve for the time evolution of correlation and
response functions, an approach which explicitly reflects the dynamic
character of the glassy state. Mostly because of its technical simplicity,
an alternative (but equivalent) approach is based on a replica theory\cite%
{Mon95,MP991}. Even though this approach does not allow one to
 calculate  the
complete time evolution, long time correlations as well as stationary
response functions can be determined which are in agreement with the
explicit dynamic theory. We will use the replica approach because of its
relative simplicity.

The central quantity of the replica theory to self generated glassiness is
the replicated partition function\cite{Mon95,MP991} 
\begin{equation}
Z\left( m\right) =\int D^{m}\phi e^{-\sum_{\alpha }^{m}H\left[ \phi _{\alpha
}\right] \ +g\int \frac{dx}{2m}\sum_{\alpha ,\beta }^{m}\phi _{\alpha
}\left( \mathbf{x}\right) \phi _{\beta }\left( \mathbf{x}\right) }.
\label{partzm}
\end{equation}%
$Z\left( m\right) $ must be analyzed in the limit $m\rightarrow 1$ and $%
g\rightarrow 0$ which is taken at the end of the calculation. One can
interpret $Z\left( m\right) $ as the result of a quenched average over
random field configurations with infinitesimal disorder strength $g$ and a
distribution function which is determined by the partition function of the
system itself. A similar approach (however for finite $g$) was proposed by
Deam and Edwards\cite{ED75} in the theory of the vulcanization transition
where the distribution of random cross-links of a polymer melt is also
determined by the actual partition function of the polymer system itself\cite%
{Paul}. \ Then, just as  in the present approach, the number of replicas, $m$%
, is taken to the limit $m\rightarrow 1$, as opposed to the usual $%
m\rightarrow 0$ limit used for systems with averaging over  "white" \
quenched disorder distributions. In the vulcanization problem, $g$ is
related to the cross link density and, in distinction to the theory
presented here, remains finite. \ Physically, the analysis of $Z\left(
m\right) $ addresses whether the system drives itself \ ($m\rightarrow 1$) \
into a glassy state if exposed to an infinitesimally weak randomness ($%
g\rightarrow 0$). It can be shown\cite{WSW02} that this approach is
equivalent to the solution of the dynamic equations for the correlation and
response function\cite{CK93} of the system that takes aging and long
term memory effects into account.

$F\left( m\right) =-\frac{T}{m}\log Z\left( m\right) $ determines the
configurational entropy, $S_{\mathrm{c}}$, and the averaged free energy of
systems trapped in metastable states, $\widetilde{F}$, via\cite%
{Mon95,MP991,SWW00} 
\begin{eqnarray}
\widetilde{F} &=&\lim_{g\rightarrow 0}\left. \frac{\partial \left( mF\left(
m\right) \right) }{\partial m}\right\vert _{m\rightarrow 1}  \nonumber \\
S_{c} &=&\lim_{g\rightarrow 0}\left. \frac{m^{2}}{T}\frac{\partial F\left(
m\right) }{\partial m}\right\vert _{m\rightarrow 1}.  \label{sc}
\end{eqnarray}
The equilibrium free energy, $F=-T\log Z_{\mathrm{eq}}$ can be shown to be $%
F=\widetilde{F}-TS_{\mathrm{c}}$. This makes evident that $TS_{\mathrm{c}}$
is indeed a part of the free energy which cannot be realized if the system
is trapped in metastable states.

The limit $m\rightarrow 1$ is only appropriate above the Kauzmann
temperature, $T_{\mathrm{K}}$, at which $S_{\mathrm{c}}$ vanishes. Below \ $T_{%
\mathrm{K}}$ one has to perform the limit $m\rightarrow T/T_{\mathrm{eff}}$
with effective temperature $T_{\mathrm{eff}}>T$ $\ $characterizing the
spectrum of metastable states. How to determine  $T_{\mathrm{eff}}$ in case of a rapid quench is
discussed in Ref. \onlinecite{WWSW02}, whereas an interesting alternative to determine $T_{\rm eff}$ for slowly quenched systems was proposed in Ref.\onlinecite{Lopatin}.  In the present paper we restrict ourselves
to the behavior $T>T_{\mathrm{K}}$ even though our approach can be
generalized to the case with arbitrary $m<1$ easily.

\subsection{Mapping onto a local problem}

The major problem is the determination of the partition function $Z\left(
m\right) $. Even for the liquid state (i.e. $m=1$ and $g=0$ at the outset)
this is a very hard problem without known exact solution and we are forced
to use computer simulations or \ to\ develop approximate analytical
theories. In developing such an approximate theory we take advantage of the
fact that glass forming systems are often driven by strong local
correlations, as opposed to the pronounced long ranged correlations at a
second order phase transition or the critical point of the liquid-vapor
coexistence curve. This is transparent both, in the mode coupling theory of
under-cooled liquids\cite{mc} and in the self-consistent phonon approaches%
\cite{SSW85}, where a given molecule is locally caged by its environment
built of other molecules. \ 

When we call a physical system\emph{\ local }this does not necessarily imply
that all its correlation functions  infinitely rapidly decay in space. In
the language of many body theory it only implies that the irreducible self
energy $\Sigma _{ll^{\prime }}\left( \mathbf{k}\right) \simeq \Sigma
_{ll^{\prime }}$ is independent of momentum for the important range of
parameters. Here, $\Sigma _{ll^{\prime }}$ is related to the correlation
function \ $G_{ll^{\prime }}\left( \mathbf{k}\right) =\left\langle \phi _{l,%
\mathbf{k}}\phi _{l^{\prime },\mathbf{-k}}\right\rangle $ via Dyson equation 
\begin{equation}
G\left( \mathbf{k}\right) ^{-1}=G_{0}\left( \mathbf{k}\right) ^{-1}-\Sigma 
\text{ }  \label{dyson}
\end{equation}%
which is an $L\times L$ matrix in case of the equilibrium \ liquid state
theory. If we study the emergence of glassy states we have to use the
replica theory and Eq.\ref{dyson} becomes an $\left( mL\times mL\right) $
matrix equation with $G_{ll^{\prime }}^{\alpha \beta }\left( \mathbf{k}%
\right) =\left\langle \phi _{l,\mathbf{k}}^{a}\phi _{l^{\prime },\mathbf{-k}%
}^{\beta }\right\rangle $, $G_{0ll^{\prime }}^{\alpha \beta }\left( \mathbf{k%
}\right) =G_{0ll^{\prime }}\left( \mathbf{k}\right) \delta _{a\beta }$ as
well as self energy matrix $\Sigma _{ll^{\prime }}^{\alpha \beta }$.

Traditionally, the self energy is introduced because it has a comparatively
simple structure within perturbation theory. However, in the theory of
strongly correlated Fermi systems it has been recognized that the existence
of a momentum independent self energy allows conceptually new insight into
the dynamics of many body systems, describing states far from the strict
perturbative limit.\cite{dmft1,dmft2} We will adopt the main strategy of
this \emph{dynamic mean field theory }  for our problem.

We use the fact that the free energy $F\left( m\right) $ determined by Eq.%
\ref{partzm} can be written as\cite{KB70} 
\begin{equation}
F\left( m\right) =\frac{1}{2}\mathrm{tr\log }\left( G_{0}^{-1}-\Sigma
\right) +\frac{1}{2}\mathrm{tr}\left( \Sigma G\right) +\frac{1}{2}\Phi \left[
G\right],   \label{KB}
\end{equation}%
where the trace goes for each $\mathbf{k}$-point over the $\left( mL\times
mL\right) $-matrix components together with a sum over $\mathbf{k}$. The
latter can also be written as a matrix trace of real space functions $%
G\left( \mathbf{x,x}^{\prime }\right) $ etc. The functional $\Phi \left[ G%
\right] $ is well defined in terms of Feynman diagrams as the sum of
skeleton diagrams of the free energy. In what follows we will not try to
calculate $\Phi $ but merely use the fact that such a functional exists.
From the definition of $\Phi $ it follows that it determines the self energy
via a functional derivative: 
\begin{equation}
\Sigma _{ll^{\prime }}^{\alpha \beta }\left( \mathbf{x,x}^{\prime }\right) =-%
\frac{\delta \Phi \left[ G\right] }{\delta G_{l^{\prime }l}^{\beta \alpha
}\left( \mathbf{x}^{\prime }\mathbf{,x}\right) }.  \label{derivative}
\end{equation}%
Since we made the assumption that $\Sigma $ is momentum dependent, Fourier
transformation yields $\Sigma _{ll^{\prime }}^{\alpha \beta }\left( \mathbf{%
x,x}^{\prime }\right) =\Sigma _{ll^{\prime }}^{\alpha \beta }\delta \left( 
\mathbf{x-x}^{\prime }\right) $. Thus, the functional derivative, Eq.\ref%
{derivative}, vanishes if $\mathbf{x}\neq \mathbf{x}^{\prime }$, which
implies that for a local theory, $\Phi $ solely depends on the local,
momentum averaged, correlation function, 
\begin{equation}
\overline{G}_{ll^{\prime }}^{\alpha \beta }=\int \frac{d^{d}k}{\left( 2\pi
\right) ^{d}}G_{ll^{\prime }}^{\alpha \beta }\left( \mathbf{k}\right) \text{.%
}
\end{equation}%
Since all our interactions, $V\left[ \phi \right] $ are (by assumption)
local as well, we conclude that there exists a local problem with
Hamiltonian 
\begin{equation}
\mathcal{H}=\sum_{ll^{\prime },\alpha \beta }\phi _{l}^{\alpha }\mathcal{J}%
_{ll^{\prime }}^{\alpha \beta }\phi _{l^{\prime }}^{\beta
}+a_{0}^{d}\sum_{\alpha }V\left[ \phi _{l}^{a}\right] ,  \label{localham}
\end{equation}%
which has an identical functional $\Phi \left[ \mathcal{G}\right] $ of its
own \ correlation function \ $\mathcal{G}$, which is also a $\left( mL\times
mL\right) $-matrix but does not depend on position or momentum. $a_{0}$ is a
typical microscopic length scale, for example a hard core diameter and needs
to be specified for each system. It results from the fact that a sensible local theory can only be formed after some appropriate discretization
\begin{equation}
\int d^dx V(\phi_l(x)) \rightarrow a_0^d \sum_i V(\phi_l(x_i))\, .
\end{equation} 
Even though, $\mathcal{H}$ has no spatial
structure anymore, the perturbation theory up to arbitrary order is the same
for both systems. This holds for an arbitrary choice of the so called Weiss
field, $\mathcal{J}$.

Following Ref.\onlinecite{dmft2}  we use the freedom to chose $\mathcal{J}$  in
order to guarantee that $\mathcal{G}=\overline{G}$. This implies that not
only the functional $\Phi $ but also its argument are the same for the
actual physical system and the auxiliary local one. It then follows that the
self energy of \ the original system $\Sigma _{ll^{\prime }}^{\alpha \beta }$
is, up to a trivial prefactor which results from the above discretization, 
equal to the self energy  of the auxiliary
system, $a_{0}^{d}\Sigma _{ll^{\prime }}^{\alpha \beta }$. The prefactor $%
a_{0}$ can for example be determined by comparing the leading order in
perturbation theory of the two problems, Eq.\ref{hamgen} and Eq.\ref%
{localham}. Then we solely need to solve the much simpler problem, $\mathcal{%
H}$, and determine for an assumed Weiss field $\mathcal{J}$ $\ $the local
self energy as well as the local propagator related by 
\begin{equation}
\overline{G}=\left( \mathcal{J}-a_{0}^{d}\Sigma \right) ^{-1}\text{. }
\label{dys1}
\end{equation}%
We made the right choice for $\mathcal{J}$ if simultaneously  to Eq.\ref%
{dys1} it is true that self energy and averaged correlation function are related
by: 
\begin{equation}
\overline{G}=\int \frac{d^{d}k}{\left( 2\pi \right) ^{d}}\left( G_{0}\left( 
\mathbf{k}\right) ^{-1}-\Sigma \right) ^{-1}.  \label{dys2}
\end{equation}%
If \ this second equation is not fulfilled we need to improve the Weiss
field $\mathcal{J}$ until Eq.\ref{dys1} and Eq.\ref{dys2} hold
simultaneously, posing a self consistency problem. This is the most
consistent way to determine the physical correlation functions under the
assumption of a momentum independent self energy. This approach has been
applied to a large class of problems in the field of strongly correlated
Fermi systems, and promises, as we argue here, to be very useful in rather
different contexts. The major technical task of the DMFT is to solve $%
\mathcal{H}$ for given $\mathcal{J}$. This will be done for the specific
choice of a replica symmetric correlation function 
\begin{equation}
G_{ll^{\prime }}^{\alpha \beta }\left( q\right) =K_{ll^{\prime }}\left(
q\right) \delta _{\alpha \beta }+F_{ll^{\prime }}\left( q\right) ,
\end{equation}%
which implies a similar form for the self energy 
\begin{equation}
\Sigma ^{\alpha \beta }=\Sigma _{ll^{\prime }}^{K}\delta _{\alpha \beta
}+\Sigma _{ll^{\prime }}^{F}.
\end{equation}%
Below we will present a general approach to study the stability of this
choice \ and analyze it for a specific example.

\subsection{Example: the Brazovskii model of micro-phase separation}

We are now in the position to apply our approach to a specific physical
system. We consider a one component system ($L=1$) governed by the three
dimensional Brazovskii model\cite{Brazovskii}, 
\begin{equation}
H=\frac{1}{2}\int d^{3}x\left( \varepsilon _{0}^{2}q_{0}^{2}\phi ^{2}+\frac{u%
}{2}\phi ^{4}+\frac{q_{0}^{-2}}{4}\left( \left[ \nabla ^{2}+q_{0}^{2}\right]
\phi \right) ^{2}\right)   \label{braz}
\end{equation}%
which has a broad range of applicability in systems with micro-phase
separation like the theory of micro-emulsions\cite%
{Stillinger83,Deem94,WWSW02}, block copolymers\cite{Leibler,Fredrickson} or
even doped transition metal oxides\cite{EK93,SW00,SWW00}. In Refs.\onlinecite{KT89,Kivelson}
 it was argued that it might be used as a simple continuum model for glass
forming liquids. \ The bare correlation function follows from Eq.\ref{braz} 
\begin{eqnarray}
G_{0}\left( q\right)  &=&\frac{1}{\varepsilon _{0}^{2}q_{0}^{2}+\ \frac{1}{%
4q_{0}^{2}}\left( q^{2}-q_{0}^{2}\right) ^{2}}  \nonumber \\
&\simeq &\frac{1}{\varepsilon _{0}^{2}q_{0}^{2}+\ \left( q-q_{0}\right) ^{2}}%
.
\end{eqnarray}%
The interesting behavior of the  Brazovskii model arises \ from the large
phase space of low energy fluctuations which is evident from the gradient
term, $\nabla ^{2}+q_{0}^{2}$, in the Hamiltonian. All fluctuations with
momenta $\left\vert \mathbf{q}\right\vert =q_{0}$\ can, independent of the
direction of $\mathbf{q}$, be excited most easily. The wave number $q_{0}$
is related to various physical quantities in all these different systems. In
microemulsions, $q_{0}$ is determined by the volume fraction of amphiphilic
molecules whereas it is inversely proportional to the radius of gyration in
block copolymers and to the strength of the Coulomb interaction in doped
transition metal oxides. Clearly, the role played by the microscopic length
scale $a_{0}$ of the previous section is $q_{0}^{-1}$. \ As discussed in
Appendix A, an explicit calculation within the self consistent screening
approximation gives $a_{0}\simeq 1.7q_{0}^{-1}$. For simplicity we use
 $a_{0}=q_{0}^{-1}$, which corresponds to a rescaling of the coupling constant.

If the self energy is momentum independent the correlation function has the
form  
\begin{equation}
G\left( q\right) =\frac{1}{\left( q-q_{0}\right) ^{2}+\varepsilon
^{2}q_{0}^{2}},
\end{equation}%
with $\varepsilon ^{2}=\varepsilon _{0}^{2}-q_{0}^{-2}\Sigma >0$, which
yields 
\begin{equation}
\overline{G}=\int \frac{d^{3}q}{\left( 2\pi \right) ^{3}}G\left( q\right) =%
\frac{q_{0}}{2\pi \varepsilon }.  \label{Gav}
\end{equation}

In Refs.\onlinecite{SW00,SWW00} it was shown that a self generated glass
transition can occur and the
configurational entropy was calculated. A glassy state occurs for all finite 
$q_{0}$ and leads to $S_{\mathrm{c}}\propto Vq_{0}^{3}$, with volume $V$.
However, these results were obtained by using perturbation theory and in
particular the discussion of Ref.\onlinecite{BCK96} suggests that the self
consistent screening approximation might strongly over estimate the tendency
towards glassiness. An important question is therefore whether the glassy
state obtained in Refs.\onlinecite{SW00,SWW00} is \ generic or arises solely as
consequence of a poor approximation. \ This issue was also raised in the
recent numerical simulations in Ref.\onlinecite{Geissler}. The approach developed
in this paper is able to answer this question. In order to facilitate the
comparison with the perturbative results we summarize the results of Refs. 
\onlinecite{SW00,SWW00} in Appendix A however using an argumentation more adapted
to the  dynamic mean field approach.

\subsubsection{DMFT for the liquid state}

Ignoring glassiness for the moment, the local Hamiltonian is given by 
\begin{equation}
\mathcal{H}=\frac{1}{2}\mathcal{J}\phi ^{2}+\frac{uq_{0}^{-3}}{4}\phi ^{4}.
\end{equation}%
which leads to the partition sum $\mathcal{Z}=\int_{-\infty }^{\infty }d\phi
e^{-\mathcal{H}}$ and correlation function 
\begin{equation}
\overline{G}=\frac{\int_{-\infty }^{\infty }d\phi \phi ^{2}e^{-\mathcal{H}}}{%
\mathcal{Z}}.  \label{Gacloc}
\end{equation}%
$\mathcal{Z}$ and $\overline{G}$ are elementary integrals that can be
expressed in terms of elliptic functions. The Weiss field which leads to the
correct propagator can be obtained as $\ \mathcal{J}=\frac{\varepsilon
_{0}^{2}+2\pi \varepsilon -\varepsilon ^{2}}{q_{0}}$ which, together with
Eqs.\ref{Gav} and \ref{Gacloc}, leads to a single nonlinear algebraic
equation for $\varepsilon $. The solution can be obtained easily. Results
for $\varepsilon _{0}^{2}=-4$ (strong segregation limit) as well as $%
\varepsilon _{0}^{2}=-1$ (weaker segregation limit) are shown in Fig.1 and
are compared with the corresponding behavior as obtained using a Hartree
theory which gives $\varepsilon _{\mathrm{Hartree}}=-\frac{u/q_{0}}{%
\varepsilon _{0}^{2}2\pi }$. \ Leading order perturbation theory (i.e.
Hartree theory) does give the proper small $u$ behavior, however strong deviations
occur  in case of the weak segregation limit (small $\left\vert
\varepsilon _{0}^{2}\right\vert $).

\begin{figure}[tbp]
\includegraphics[width=3.0in,angle=-90]{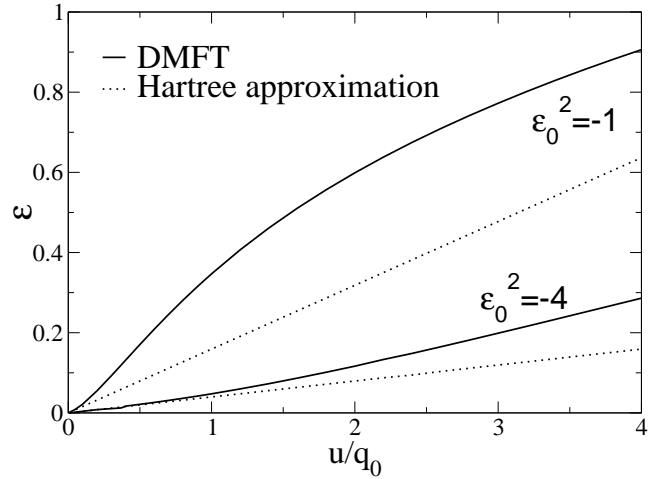}
\caption{Dimensionless inverse correlation length of the equilibrium liquid
state as function of the coupling constant (scaled by the modulation wave
number $q_0$ for different segregation strength $\protect\epsilon_0^2$. The
DMFT results (solid lines) are compared with the leading order Hartree
approximation (dashed lines). Strong deviations for small dimensionless
coupling constants $u/q_0$ occur in the limit of weak segregation strength.
Note, the results are obtained by ignoring the possibility of a fluctuation
induced first order transition which will become relevant if $|\protect%
\epsilon_0^2|> c (u/q_0)^(2/3)$ with $c$ a constant of order unity.}
\end{figure}

In our treatment of the equilibrium behavior we have made the assumption
that no phase transition to a state with long range order of $\left\langle
\phi \left( \mathbf{x}\right) \right\rangle $ takes place. This is
consistent with the Hartree theory which excludes a diverging correlation
length for finite temperatures, and thus, excludes the possibility of a
second order phase transition. However this assertion contrasts with the
result found by Brazovskii\cite{Brazovskii} who showed that the model, Eq.%
\ref{braz}, undergoes a fluctuation induced first order transition to a
modulated state with order parameter \ 
\begin{equation}
\left\langle \phi \left( \mathbf{x}\right) \right\rangle =\phi _{0}\cos
\left( \mathbf{q}_{0}\cdot \mathbf{x}\right) ,  \label{BROP}
\end{equation}%
where $\left\vert \mathbf{q}_{0}\right\vert =q_{0}$ and arbitrary but fixed
direction. For small coupling constant this transition occurs if $\left\vert
\varepsilon _{0}^{2}\right\vert >\frac{3}{\left( 8\pi \right) ^{2/3}}\left( 
\frac{u}{q_{0}}\right) ^{2/3}$. The criterion for the glass transition
discussed in this paper is very similar to the parameter values leading to
the fluctuation
induced first order transition. We have not analyzed both transitions within
an identical theoretical approach but find in Appendix A that perturbation
theory yields for the occurrence of the glass transition $\left\vert
\varepsilon _{0}^{2}\right\vert \propto \left( \frac{u}{q_{0}}\right) ^{3/5}$%
. In what follows we will always assume that the glass transition occurs 
\emph{instead }of the fluctuation induced first order transition. In the
laboratory this will likely be \ a kinetic issue, that needs a nucleation
theory for quantitative predictions. The results of Ref.\onlinecite{Hohenberg}
where a nucleation theory of the Brazovskii transition was developed
demonstrate that indeed the nucleation kinetics of the order parameter, Eq.%
\ref{BROP} is very complex, i.e. under-cooling should be possible.

\subsubsection{DMFT in the glassy state}

\ To use the replica theory of the glassy state we have to specify the
replica structure of the correlation function. We first choose a given
structure and discuss its stability later. In replica space we start from
the following structure of the propagators and self energies: 
\begin{eqnarray}
G_{\alpha \beta }\left( q\right)  &=&K\left( q\right) \delta _{\alpha \beta
}+F\left( q\right)   \nonumber \\
\Sigma _{\alpha \beta } &=&\Sigma _{K}\delta _{\alpha \beta }+\Sigma _{F}.
\label{replicaGG}
\end{eqnarray}%
Inverting the matrix Dyson equation in replica space leads to: 
\begin{eqnarray}
K\left( q\right)  &=&\frac{1}{\left( q-q_{0}\right) ^{2}+\kappa ^{2}q_{0}^{2}%
}   \\
mF\left( q\right)  &=&\frac{1}{\left( q-q_{0}\right) ^{2}+\varepsilon
^{2}q_{0}^{2}}-\frac{1}{\left( q-q_{0}\right) ^{2}+\kappa ^{2}q_{0}^{2}}
\nonumber
\end{eqnarray}
with the convenient definitions: 
\begin{eqnarray}
\kappa ^{2}q_{0}^{2} &=&\varepsilon _{0}^{2}q_{0}^{2}-\Sigma _{K}\left(
q\right)    \\
\varepsilon ^{2}q_{0}^{2} &=&\varepsilon _{0}^{2}q_{0}^{2}-\Sigma _{K}\left(
q\right) -m\Sigma _{F}\left( q\right) . \nonumber
\end{eqnarray}

The diagonal elements $K\left( q\right) +F\left( q\right) $ can be
interpreted as the equilibrium, liquid state correlation function and is
only determined by $\varepsilon $ which can be related to the liquid state
correlation length $\xi \simeq \frac{1}{\varepsilon q_{0}}$. On the other
hand $F\left( q\right) =\lim_{t\rightarrow \infty }\lim_{t^{\prime
}\rightarrow \infty }\left\langle \phi _{q}\left( t\right) \phi _{-q}\left(
t+t^{\prime }\right) \right\rangle $ characterizes long time correlations in
analogy to the Edwards-Anderson parameter. \ Clearly, if $\Sigma _{F}\neq 0$
then $\kappa >\varepsilon $ $\ $and $F\left( q\right) >0$. \ Depending of
whether one considers $q$ values close to or away from $q_{0}$, $F\left(
q\right) $ is governed by the correlation length $\xi $ or the Lindemann
length of the glass $\lambda _{0}=q_{0}^{-1}\left( \kappa ^{2}-\varepsilon
^{2}\right) ^{-1}$, respectively. The physical significance of \ $\lambda
_{0}$ is the length scale over which defects and imperfections of \ an
crystalline state can wander after long time was discussed in Ref.\cite%
{SWW00}. Finally, the correlation function $K\left( q\right) $ (which is
solely determined by the short length $\frac{1}{\kappa q_{0}}<\xi $) is the
response function of a local perturbation. Obviously, any response of the
glassy system is confined to very small length scales even though the
instantaneous correlation length can be considerable. This is a clear
reflection of the violation of the fluctuation dissipation relation within
the replica approach. Averaging these functions over momenta gives: 
\begin{eqnarray}
\overline{K} &=&\int \frac{d^{3}q}{\left( 2\pi \right) ^{3}}K\left( q\right)
=\frac{q_{0}}{2\pi \kappa },  \nonumber \\
\overline{F} &=&\int \frac{d^{3}q}{\left( 2\pi \right) ^{3}}F\left( q\right)
=\frac{q_{0}}{2\pi m}\left( \frac{1}{\varepsilon }-\frac{1}{\kappa }\right) .
\label{av}
\end{eqnarray}

The \ auxiliary local Hamiltonian is 
\begin{equation}
\mathcal{H}=\frac{1}{2}\sum_{ab}\mathcal{J}_{ab}\phi _{a}\phi _{b}+\frac{%
uq_{0}^{-3}}{4}\sum_{a}\phi _{a}^{4}.
\end{equation}%
Within DMFT we then find for its correlation function 
\begin{equation}
\mathcal{G}=\overline{K}\delta _{ab}+\overline{F}.
\end{equation}%
In addition, the Weiss field is given by: 
\begin{equation}
\mathcal{J}_{\alpha \beta }=\mathcal{J}\delta _{\alpha \beta }-\mathcal{C}.
\end{equation}%
These relations together with Eq.\ref{av} can be used to express the Weiss
fields in terms of the $\varepsilon $ and $\kappa $: 
\begin{eqnarray}
\mathcal{J} &=&\frac{2\pi \kappa +\varepsilon _{0}^{2}-\kappa ^{2}}{q_{0}} 
\nonumber \\
\mathcal{C} &=&\frac{2\pi \left( \kappa -\varepsilon \right) +\varepsilon
^{2}-\kappa ^{2}}{mq_{0}}.
\label{Jinte}
\end{eqnarray}%
Thus, we have to determine $\overline{K}$ and $\overline{F}$ for given $%
\mathcal{J}$ and $\mathcal{C}$ and make sure that the latter are chosen such
that Eqs.\ref{av}  and \ref{Jinte} are fulfilled.

The partition sum of the local problem is given by 
\begin{equation}
\mathcal{Z}\left( m\right) =\int d^{m}\phi e^{-\sum_{\alpha =1}^{m}\mathcal{H%
}_{0}\left[ \phi _{\alpha }\right] +\frac{1}{2}\mathcal{C}\left(
\sum_{\alpha }\phi _{\alpha }\right) ^{2}},
\end{equation}
where $\mathcal{H}_{\mathrm{0}}\left[ \phi _{\alpha }\right] =\frac{1}{2}%
\mathcal{J}\phi _{a}^{2}+\frac{uq_{0}^{-3}}{4}\phi _{a}^{4}$. \ $d^{m}\phi $
refers to the fact that $\phi $ is an $m$-component vector and the integral
goes over an $m$-dimensional space with arbitrary $m$. The coupling between
different replicas can be eliminated by performing a Hubbard-Stratonovich
transformation, which leads to 
\begin{equation}
\mathcal{Z}\left( m\right) =\int \frac{d\lambda }{\sqrt{2\pi \mathcal{C}}}%
e^{-\frac{\lambda ^{2}}{2\mathcal{C}}}\Omega \left( \lambda \right) ^{m}
\end{equation}
where 
\begin{equation}
\Omega \left( \lambda \right) =\int d\phi e^{-\left( \mathcal{H}_{0}\left[
\phi \right] +\lambda \phi \right) }
\end{equation}
is the equilibrium partition function however in an external\ field $\lambda 
$, with Gaussian distribution function.

In order to determine the propagators of the local problem, we consider the
sum of the diagonal elements 
\begin{equation}
\sum_{a}\overline{G}_{aa}=m\left( \overline{K}+\overline{F}\right) \ 
\end{equation}
which is equal to $-2\frac{\partial \log \mathcal{Z}\left( m\right) }{%
\partial \mathcal{J}}$, yielding 
\begin{equation}
\overline{K}+\overline{F}=-2\int \frac{d\lambda e^{-\frac{\lambda ^{2}}{2%
\mathcal{C}}}\Omega \left( \lambda \right) ^{m-1}}{\mathcal{Z}\left(
m\right) \sqrt{2\pi \mathcal{C}}}\frac{\partial }{\partial \mathcal{J}}%
\Omega \left( \lambda \right) .
\end{equation}
The derivative with respect to $\mathcal{J}$ leads to 
\begin{equation}
\frac{\partial }{\partial \mathcal{J}}\Omega \left( \lambda \right) =-\frac{1%
}{2}\int d\phi e^{-\left( \frac{1}{2}\mathcal{J}\phi ^{2}+\frac{uq_{0}^{-3}}{%
4}\phi ^{4}+\lambda \phi \right) }\phi ^{2}
\end{equation}
which gives the final expression for the diagonal element of the replica
correlation function 
\begin{equation}
\overline{K}+\overline{F}=\frac{\int \frac{d\lambda }{\sqrt{2\pi \mathcal{C}}%
}e^{-\frac{\lambda ^{2}}{2\mathcal{C}}}\Omega \left( \lambda \right)
^{m}\left\langle \phi ^{2}\right\rangle _{\lambda }}{\mathcal{Z}\left(
m\right) }  \label{diag}
\end{equation}
with 
\begin{equation}
\left\langle \phi ^{2}\right\rangle _{\lambda }=\frac{\int d\phi e^{-\left( 
\frac{1}{2}\mathcal{J}\phi ^{2}+\frac{uq_{0}^{-3}}{4}\phi ^{4}+\lambda \phi
\right) }\phi ^{2}}{\Omega \left( \lambda \right) }.
\end{equation}

In addition we also need to determine the off diagonal elements in replica
space of the correlation function. We use 
\begin{equation}
\sum_{a,b}\overline{G}_{ab}=m\overline{K}+m^{2}\overline{F}\ 
\end{equation}%
which equals to $2\frac{\partial \log \mathcal{Z}\left( m\right) }{\partial 
\mathcal{C}}$ and obtain an equation which can be used to determine the off
diagonal elements $\overline{F}$ 
\begin{equation}
\overline{K}+m\overline{F}=-\frac{\int \frac{d\lambda }{\sqrt{2\pi \mathcal{C%
}}}e^{-\frac{\lambda ^{2}}{2\mathcal{C}}}\Omega \left( \lambda \right) ^{m}%
\mathcal{C}^{-1}\left\langle \phi \right\rangle _{\lambda }\lambda .}{%
\mathcal{Z}\left( m\right) }  \label{offdb}
\end{equation}%
Evidently, Eqs.\ref{diag} and \ref{offdb} are only independent\ if $m$
differs from $1$.

We next analyze these equations for small but finite $m-1$. This can be done
by expanding Eqs.\ref{diag} and \ref{offdb} into a Taylor series for small $%
m-1$ and comparing order by order. First we consider the zeroth order term
and find that both equations yield for $m=1$, the results for the liquid
state which determine $\varepsilon $. The next step is to consider the first
corrections linear in $m-1$. This allows us to check whether there are
nontrivial solutions for $\kappa >\varepsilon $ and thus for the off
diagonal self energy and long time correlation function.

The difference between Eqs.\ref{diag} and \ref{offdb} gives

\begin{equation}
\overline{F}=\frac{\int \frac{d\lambda }{\sqrt{2\pi \mathcal{C}}}e^{-\frac{%
\lambda ^{2}}{2\mathcal{C}}}\Omega \left( \lambda \right) ^{m}\left(
\left\langle \phi ^{2}\right\rangle _{\lambda }+\mathcal{C}^{-1}\left\langle
\phi \right\rangle _{\lambda }\lambda \right) }{\left( 1-m\right) \mathcal{Z}%
\left( m\right) }.
\end{equation}
Expanding the numerator for $m$ close to $1$ gives 
\begin{equation}
\overline{F}=\frac{\delta I\ }{\mathcal{Z}\left( m=1\right) }
\end{equation}
with 
\begin{equation}
\delta I=-\int \frac{d\lambda \left( \mathcal{C}^{-1}\left\langle \phi
\right\rangle _{\lambda }\lambda +\left\langle \phi ^{2}\right\rangle
_{\lambda }\right) e^{-\frac{\lambda ^{2}}{2\mathcal{C}}}\Omega \left(
\lambda \right) \log \Omega \left( \lambda \right) }{\sqrt{2\pi \mathcal{C}}}%
.  \label{dI}
\end{equation}

As expected, it follows that $\overline{F}=0$ if $\kappa =\varepsilon $,
where $\mathcal{C}=0$. This can be seen from the expansion $\left\langle
\phi \right\rangle _{\lambda }\simeq -\left\langle \phi ^{2}\right\rangle
_{\lambda }\lambda $, valid for small $\lambda $ and by substituting $\mu
=\lambda /\sqrt{\mathcal{C}}$. It follows that
\begin{equation}
\overline{F}\left( \kappa =\varepsilon \right) \ \propto \int \frac{d\mu }{%
\sqrt{2\pi }}e^{-\frac{\mu ^{2}}{2}}\left( 1-\mu ^{2}\right) =0.
\end{equation}

\begin{figure}[tbp]
\includegraphics[width=3.0in,angle=-90]{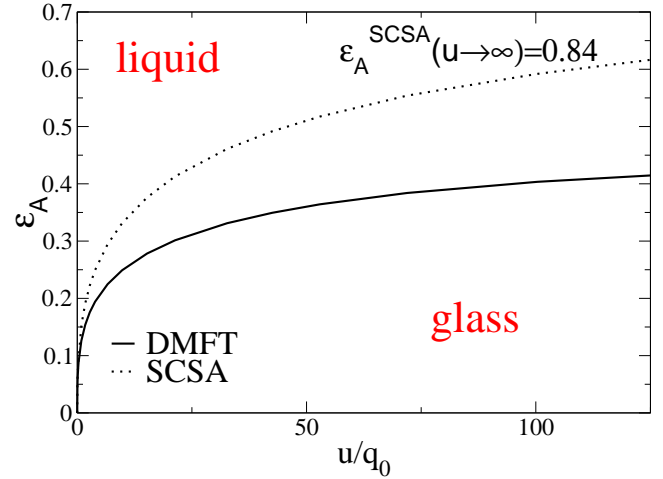}
\caption{Dimensionless inverse correlation length as function of the coupling
constant at the dynamical glass transition $\protect\epsilon_A$ obtained
within the DMFT, in comparison with the result of the self consistent
screening approximation. For large $u$ a universal value of the correlation
length (in units of $q_0^{-1}$ for glassiness results, whereas the critical
value of the spatial correlations to reach a glassy state increases
considerably in the weak coupling limit. }
\end{figure}

In Fig.2 we show our results for the dimensionless inverse correlation
length $\varepsilon _{\mathrm{A}}=\varepsilon \left( u_{\mathrm{A}}\right) $
at the onset of glassiness. We have to distinguish the two different regimes
of strong and weak phase separation, characterized by $\left\vert
\varepsilon _{0}^{2}\right\vert >1$ and $\left\vert \varepsilon
_{0}^{2}\right\vert <1$, respectively. \ In the case of strong phase
separation we find a universal \ value $\varepsilon _{\mathrm{A}}\simeq 0.45$%
. The universality of this result is in qualitative agreement with results
obtained with perturbation theory which also gives such a generic value
which is however numerically larger ($\varepsilon _{\mathrm{A}}\simeq 0.8385$%
). In addition we obtain that at the transition the ratio $\kappa _{\mathrm{A%
}}/\varepsilon _{\mathrm{A}}$ $\simeq 3.5$, which is close to the result $%
\kappa _{\mathrm{A}}/\varepsilon _{\mathrm{A}}=3.1$, obtained within
perturbation theory.The onset of glassiness in the model Eq.\ref{braz}
occurs in the strong phase separation limit and, for given $q_{0}$, is
uniquely determined by the value of the correlation length $\xi =\left(
\varepsilon q_{0}\right) ^{-1}$, and not explicitly dependent on $u$ or $T$.
\ The exact solution of the dynamic mean field problem leads to the
interesting result that this value for $\xi $ is larger than what follows
within perturbation theory. For small $\frac{u}{q_{0}}$, the result of the
perturbation theory is $\varepsilon _{\mathrm{A}}\simeq 0.42\left( \frac{u}{%
q_{0}}\right) ^{2/5}$. In this limit the DMFT gives the interesting result
that $\varepsilon _{\mathrm{A}}$ is larger than the value obtained within
perturbation theory. The effect is small and
numerically hard to identify. It implies that the criterion for glassiness
as obtained within DMFT becomes very close to the one found by Brazovskii
for the fluctuation induced first order transition. Most important \ however
is that we do indeed find that there is a nontrivial solution $\kappa
>\varepsilon $. The emergence of a self generated glassy state is therfore 
not a consequence of the perturbative solution used in Refs.\onlinecite{SW00,SWW00}%
.  Our calculation is performed for a temperature $T=1$. We can
reintroduce the temperature into the calculation by substituting $%
u\rightarrow uT$. A critical value for $u$ leads to a temperature $T_{%
\mathrm{A}}$ (for fixed $u$) where within mean field theory an exponential
number of metastable states emerges. 

\begin{figure}[tbp]
\includegraphics[width=3.0in,angle=-90]{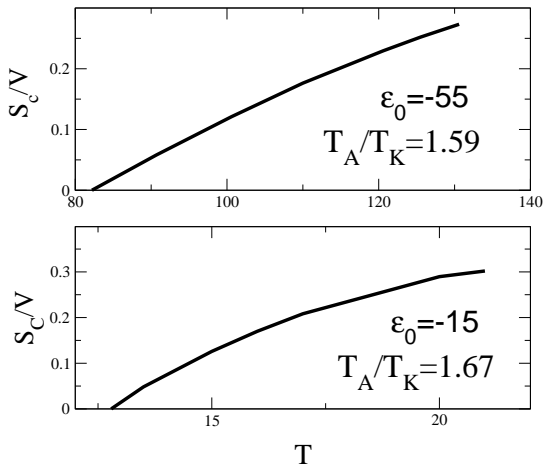}
\caption{Configurational entropy density $S_c/V$ (with system volume $V$) as
function of temperature for two different values of the segregation strength $%
\protect\epsilon_0^2$. }
\end{figure}

Once the correlation functions are determined we can use the fact that the
functional $\Phi \left[ G\right] $ is the same for the local problem as well
as the original one to obtain 
\begin{equation}
F\left( m\right) =\mathcal{F}\left( m\right) +\frac{T}{2m}\left( \mathrm{%
tr\log }G^{-1}-a_0^{-d}\mathrm{Tr\log }\overline{G}^{-1}\right) .
\end{equation}%
Here $\mathrm{Tr}$ refers to the trace over replicas, but \ does not include
the momentum integration, as opposed to $\mathrm{tr}$ which corresponds to a
trace with respect to all degrees of freedom. Finally, $\mathcal{F}\left(
m\right) =-\frac{T}{m}\log \mathcal{Z}\left( m\right) $ is the counter part
of $F\left( m\right) $ for the local problem. Using Eq.\ref{sc} an expression
for the configurational entropy follows: 
\begin{eqnarray}
\frac{S_{c}}{V} &=&-\frac{1}{2}\sum_{q}[\log \mathbf{(}1-\frac{F(q)}{G(q)}\mathbf{)+}%
\frac{F(q)}{G(q)}]-\frac{a_0^{-d}}{2}\log \left( \frac{\kappa }{\varepsilon }%
\right)   \nonumber \\
&&+\frac{a_0^{-d}}{2}(1-\frac{\varepsilon }{\kappa }) -a_0^{-d}\left. \frac{%
\partial \log \mathcal{Z}\left( m\right) /m}{\partial m}\right\vert
_{m\rightarrow 1},
\end{eqnarray}%
where $G(q)=K(q)+F(q)$ and 
\begin{equation}
\left. \frac{\partial \mathcal{Z}\left( m\right) }{\partial m}\right\vert
_{m\rightarrow 1}=\int \frac{d\lambda }{\sqrt{2\pi \mathcal{C}}}e^{-\frac{%
\lambda ^{2}}{2\mathcal{C}}}\Omega \left( \lambda \right) \log \Omega \left(
\lambda \right) .
\end{equation}%
As expected, in the limit $\kappa \rightarrow \varepsilon $ without glassy
long time correlations, it follows\ $\mathcal{Z}\left( 1\right) =\Omega
\left( 0\right) $ as well as $\left. \frac{\partial \mathcal{Z}\left(
m\right) }{\partial m}\right\vert _{m\rightarrow 1}=\Omega \left( 0\right)
\log \Omega \left( 0\right) $ and $F(q)$ goes to zero,\ leading to $%
S_{c}\left( \kappa =\varepsilon \right) =0$. \ The configurational entropy
is finite only for nontrivial solutions $\kappa >\varepsilon $. The
temperature, $T_{\mathrm{A}}$, where this happens for the first time is
equal to the dynamic transition temperature of the system.  The result for $%
S_{c}\left( T\right) $ also enables us to determine the Kauzmann
temperature, $T_{\mathrm{K}}<T_{\mathrm{A}}$, where $S_{\rm c}$ vanishes.  
In Fig.3, we show 
$S_{\rm c}(T)$
for two  different $\left\vert \varepsilon _{0}^{2}\right\vert .$ The
configurational entropy vanishes as $T_{\mathrm{K}}$ is approached: $%
S_{c}\propto T-T_{\mathrm{K}}$. The regime $(T_{\rm A}-T_{\rm K})/T_{\rm A}$ between the onset of slow dynamics and entropy crisis increases for decreasing strength of the phase separation. 



\subsection{ Stability of the solution}

An important simplification of our approach resulted from the simple form,
Eq.\ref{replicaGG}, of the correlation function in replica space. All
diagonal elements as well as all off diagonal elements are assumed to be
identical. Whether this assumption is indeed stable can be addressed by
evaluating the eigenvalues of the stability matrix 
\begin{equation}
\widehat{H}_{\alpha \beta ;\gamma \delta }=\frac{\delta ^{2}F}{\delta
G^{\alpha b}\left( q\right) \delta G^{\gamma \delta }\left( q^{\prime
}\right) }.
\end{equation}%
If there are negative eigenvalues of $\widehat{H}$ our assumption for the
replica structure is unstable. Following Ref.\onlinecite{dAT78} we find that the
lowest eigenvalue with respect to the replica indices is determined by the
lowest eigenvalue of the matrix\cite{WSW02} 
\begin{equation}
\widehat{h}_{\mathbf{q,q}^{\prime }}=\delta \left( \mathbf{q-q}^{\prime
}\right) K^{-2}\left( q\right) +v_{0}  \label{repl}
\end{equation}%
in momentum space, where $v_{0}=\frac{\delta ^{2}\Phi }{\delta \overline{G}%
^{\alpha \beta }\delta \overline{G}^{\alpha \beta }}-2\frac{\delta ^{2}\Phi 
}{\delta \overline{G}^{\alpha \beta }\delta \overline{G}^{\alpha \delta }}+%
\frac{\delta ^{2}\Phi }{\delta \overline{G}^{\alpha \beta }\delta \overline{G%
}^{\gamma \delta }}$ with distinct $\alpha ,\beta ,\gamma $ and $\delta $.
In deriving this result we started from Eq.\ref{KB} but used the fact that
the functional $\Phi $ only depends on the momentum averaged correlation
function, such that $v_{0}$ becomes momentum independent. Eq.\ref{repl} is
similar to the Schr\"{o}dinger equation in momentum space of a single
particle with bare Hamiltonian $K^{-2}\left( q\right) $ and local potential $%
v_{0}$. The lowest eigenvalue, $E$, of this problem is given by 
\begin{equation}
1=v_{0}\int \frac{d^{3}q}{\left( 2\pi \right) ^{3}}\left( K^{-2}\left(
q\right) +E\right) ^{-1}.
\end{equation}%
This equation can be analyzed if we find a way to calculate $v_{0}$ which,
due to Eq.\ref{derivative}, is determined by the first derivative of the
self energy with respect to the correlation function $\frac{\delta \Sigma
_{\alpha \beta }}{\delta \overline{G}_{\gamma \delta }}$. This derivative
can be evaluated by following closely Baym and Kadanoff\cite{KB70}. First we
add to the local Hamiltonian an additional term $-\sum_{\alpha \beta
}U_{\alpha \beta }\phi _{a}\phi _{\beta }$ and analyze all correlation
functions for finite $U$. At the end we \ will take the limit $U\rightarrow
0 $ and the correlation function has the simple structure Eq.\ref{repl}. For
finite $U$, the correlation function is determined from the function $%
\mathcal{Z}\left[ U\right] $: 
\begin{equation}
\overline{G}_{\alpha \beta }\left[ U\right] =\left\langle \phi _{\alpha
}\phi _{\beta }\right\rangle =-\frac{\delta \log \mathcal{Z}\left[ U\right] 
}{\delta U_{\alpha \beta }}
\end{equation}%
The self energy is a functional of only $\overline{G}\left[ U\right] $ and
not of $U$ explicitly such that 
\begin{equation}
\frac{\delta \Sigma _{\alpha \beta }}{\delta U_{\gamma \delta }}=\sum_{\mu
\nu }\frac{\delta \Sigma _{\alpha \beta }}{\delta \overline{G}_{\mu \nu }}%
\frac{\delta \overline{G}_{\mu \nu }}{\delta U_{\gamma \delta }}
\end{equation}%
If we furthermore introduce 
\begin{eqnarray}
L_{\alpha \beta ;\gamma \delta } &=&\frac{\delta G_{\alpha \gamma }}{\delta
U_{\delta \beta }}  \nonumber \\
&=&\left\langle \phi _{\alpha }\phi _{\beta }\phi _{\gamma }\phi _{\delta
}\right\rangle -\left\langle \phi _{\alpha }\phi _{\gamma }\right\rangle
\left\langle \phi _{\delta }\phi _{\beta }\right\rangle
\end{eqnarray}%
which can be evaluated explicitly once $U=0$, one finds 
\begin{equation}
\sum_{\mu \nu }\left( G_{\alpha \mu }^{-1}G_{\nu \gamma }^{-1}-\frac{\delta
\Sigma _{\mu \nu }}{\delta \overline{G}_{\alpha \gamma }}\right) L_{\alpha
\beta ;\gamma \delta }=\delta _{\mu \delta }\delta _{\beta \nu }
\end{equation}%
which determines $\frac{\delta \Sigma _{\mu \nu }}{\delta \overline{G}%
_{\alpha \gamma }}$ and thus $v_{0}$.

Applying this approach to the Brazovskii model we find that the replica
structure is marginally stable at the temperature $T_{\mathrm{A}}$ where the
glassy state occurs for the first time. Below $T_{\mathrm{A}}$ the replica
symmetric ansatz Eq.\ref{repl} becomes unstable however it can be made
stable if the replica index $m$ does not approach $1$ anymore but rather
takes a value $m=\frac{T}{T_{\mathrm{eff}}}$ which defines the effective
temperature $T_{\mathrm{eff}}$ of the glass. This is a situation similar to
one step replica symmetry breaking with break point given by $m$.\cite{Mon95}
A detailed discussion of this issue is presented in Ref.\onlinecite{WSW02}.

Stability of the replica symmetric \ ansatz was only possible because we
consistently made the assumption of the dynamical mean field theory. Going
beyond the local approach, for example by using cluster DMFT techniques,
enables one to study whether or not nonlocal phenomena change the replica
structure of the theory.

\section{Conclusions}

Based on the recent development in the theory of strongly interacting
electron systems we developed a dynamic mean field theory for self generated
glasses in a continuum field theory model. The key assumption of our
approach, which applies to physical systems with short range higher order
virial coefficients, is that glass formation is the consequence of
predominantly local correlations. It is then possible to map the problem
onto a purely local theory with same interaction and with a Gaussian part of
the energy which is determined self consistently. \ The comparatively simple
approach can easily be applied to multi component systems. \ Most
importantly, recent developments in the cluster DMFT approach allow one  to
generalize this theory to include at least short range non-local effects. 
Owing to its local
character a generalization of the present treatment to include spatially
inhomogeneous states (corresponding with instantons) and thereby explicitly treat the mosaic formation
predicted in random first order transitions seems well within reach.

\section{Acknowledgments}

We would like to thank Giulio Biroli,  Paul Goldbart and Harry Westfahl
Jr. for helpful discussions. This research was supported by an award from
Research Corporation (J.S.), the Institute for Complex Adaptive Matter, the
Ames Laboratory, operated for the U.S. Department of Energy by Iowa State
University under Contract No. W-7405-Eng-82 (S.W. and J. S.), and the
National Science Foundation grants DMR-0096462 (G. K.) and CHE-9530680 (P.
G. W.).

\appendix

\section{ Self-consistent screening approximation}

In this appendix we summarize the solution of the Brazovskii model within
self-consistent screening approximation of Refs.\onlinecite{SW00,SWW00}. In Ref. 
\onlinecite{SW00} a complete numerical solution of the self consistent screening
approximation was developed which did not make any assumption with respect
to the momentum dependence of the self energy. \ It was found that $\Sigma
\left( q\right) $ does weakly depend on $q$ for $q<2q_{0}$ allowing to
approximate $\Sigma \left( q\right) \simeq \Sigma \left( q_{0}\right) $
momentum independent. This simplification was helpful to develop an
analytical solution of the same problem, presented in Ref.\onlinecite{SWW00}. Here
we will show that the results of Ref.\onlinecite{SWW00} can easily be re-derived by
performing a dynamic mean field theory where the local problem is solved
within perturbation theory.

It is useful to first summarize the problems which result in a perturbative
treatment of the problem, \ something that can already be seen in the liquid
state.  Consider a given Feynman diagram for the self energy of a theory
with interaction $u\phi ^{r}$ ($r=4$ in our case). The number of internal
lines is $I_{n}=\frac{1}{2}\left( nr-2\right) $, whereas the number of
internal momentum integrations of the original theory is $L_{n}=I_{n}-\left(
n-1\right) =n\left( \frac{r}{2}-1\right) $. In a local theory we have for
each running line a contribution $q_{0}^{-3}\overline{G}$ (i.e. we replace $%
G\left( q\right) $ by its momentum averaged value). Thus, we obtain for a
contribution to the self energy of order $n$
\begin{equation}
\delta \Sigma ^{\left( n\right) }\sim u^{n}\left( q_{0}^{3}\right) ^{L_{n}} 
\left[ q_{0}^{-3}\overline{G}\right] ^{I_{n}}  \label{seterns}
\end{equation}%
Using $r=4$ and $\overline{G}$ of Eq.\ref{Gav} it follows $\delta \Sigma
^{\left( n\right) }\sim \frac{uq_{0}}{\varepsilon }\left( \frac{u}{%
q_{0}\varepsilon ^{2}}\right) ^{n-1}$. Clearly, in the limit of $\varepsilon
<1$, where glassiness was found in Refs.\onlinecite{SW00,SWW00}, the
perturbation series is a sum  terms which are increasingly  larger in magnitude.e.  strongly divergent. The self consistent screening
approximation corresponds to the sum 
\begin{equation}
\Sigma _{\mathrm{SSC}}\sim -\sum_{n=1}^{\infty }\left( -1\right) ^{n}\delta
\Sigma ^{\left( n\right) }\propto -\frac{uq_{0}}{\varepsilon }\frac{1}{1+%
\frac{u}{q_{0}\varepsilon ^{2}}}.
\end{equation}%
Even though $\Sigma _{\mathrm{SSC}}=-\ q_{0}^{2}\varepsilon $ is small for
small $q_{0}\varepsilon ^{2}$, it is not clear at all whether this procedure
leads to reliable results, making evident the need to use a method not based on the summation of diagrams.

Using the replica approach for the local problem we obtain within the self consistent screening approximation for the local self energies:
\begin{eqnarray}
a_{0}^{3}\Sigma _{K} &=&-2\left( D_{K}K+D_{K}F+D_{F}K\right)  \nonumber \\
a_{0}^{3}\Sigma _{F} &=&-2D_{F}F
\end{eqnarray}%
with collective propagators 
\begin{eqnarray}
D_{K} &=&\frac{1}{q_{0}^{3}u^{-1}+\Pi _{K}},  \nonumber \\
D_{F} &=&-\frac{\Pi _{F}D_{K}^{2}}{\ 1+\Pi _{F}D_{K}}\ ,
\end{eqnarray}%
where $\Pi _{K}=K^{2}+2KF$ and $\Pi _{F}=F^{2}$. This corresponds to the
summation of bubble diagrams as summarized in detail in Ref.\onlinecite{SWW00} and follows from the functional
\begin{equation}
\Phi[G]=a_0^{-d}{\rm Tr}{\rm log}D^{-1}
\end{equation}
with $D_{ab}=D_K\delta_{ab}+D_F$.
Since we are solving a classical (time independent) local problem we do not
need to perform any momentum integrations in the evaluation of the diagrams.

Using the expressions for the momentum averaged propagator, Eq.\ref{av}, we
find for $\Sigma _{F}=\left( \kappa ^{2}-\varepsilon ^{2}\right) q_{0}^{2}$:$%
\ $ 
\begin{equation}
a_{0}^{3}\Sigma _{F}=\frac{4\pi \varepsilon }{q_{0}\left( \ 1+\varepsilon
^{2}v^{-1}\right) }\frac{\ \left( 1-\frac{\varepsilon }{\kappa }\right) ^{3}%
}{\varepsilon ^{2}v^{-1}+1-\left( 1-\frac{\varepsilon }{\kappa }\right) ^{2}}%
,\ 
\end{equation}%
where $v=\frac{1}{\left( 2\pi \right) ^{2}}uq_{0}^{2}a_{0}^{3}$. With the
choice $a_{0}=\ \frac{\pi ^{2/3}}{2^{1/3}}q_{0}^{-1}\simeq 1.7q_{0}^{-1}$
this agrees with the result of Ref.\onlinecite{SWW00}, analyzed in the strong
coupling limit $v^{-1}\rightarrow 0$. In what follows we simply use $%
a_{0}=q_{0}^{-1}$ to be able to compare the results of the self consistent
screening with the solution of the DMFT approach of this paper. In the limit 
$v^{-1}\rightarrow 0$ a solution $\kappa >\varepsilon $ exists for the first
time if $\varepsilon <\varepsilon _{\mathrm{A}}\simeq 0.84$. At the
transition $\kappa _{\mathrm{A}}\simeq 3.1\varepsilon _{\mathrm{A}}$. \ A
detailed physical discussion of these results is given in Ref.\onlinecite{SWW00}.
For weak coupling it follows $\varepsilon _{\mathrm{A}}\simeq 0.422\left( 
\frac{u}{q_{0}}\right) ^{2/5}$ with $\kappa _{\mathrm{A}}\simeq
2.19\varepsilon _{\mathrm{A}}$.


\begin{thebibliography}{99}
\bibitem{mang01} D. N. Argyriou, J. W. Lynn, R. Osborn, B. Campbell, J. F.
Mitchell, U. Ruett, H. N. Bordallo, A. Wildes, C. D. Ling, Physical Review
Letters \textbf{89}, 036401 (2002).

\bibitem{Millis96} A. J. Millis, Phys. Rev. B \textbf{53}, 8434 (1996).

\bibitem{Dagotto} E. Dagotto, T. Hotta, and A. Moreo, Physics Reports
(2001). 

\bibitem{nmr01} M.-H. Julien, F. Borsa, P. Carretta, M. Horvatic, C.
Berthier, and C. T. Lin, Phys. Rev. Lett. \textbf{83}, 604 (1999).

\bibitem{nmr02} A. W. Hunt, P. M. Singer, K. R. Thurber, and T. Imai, Phys.
Rev. Lett. \textbf{82}, 4300 (1999).

\bibitem{nmr03} N. J. Curro, P. C. Hammel, B. J. Suh, M. H\"{u}cker, B. B%
\"{u}chner, U. Ammerahl, and A. Revcolervschi, Phys. Rev. Lett. \textbf{85},
642 (2000).

\bibitem{nmr3b} J. Haase, R. Stern, C. T. Milling, C. P. Slichter, and D. G.
Hinks, Physica C \textbf{341}, 1727 (2000).

\bibitem{nmr04} N. J. Curro, Journal of Physics and Chemistry of Solids 
\textbf{63}, 2181(2002).

\bibitem{msr01} Ch. Niedermeyer, C. Bernhard, T. Blasius, A. Golnik, A.
Moodenbaugh, and J. I. Budnik, Phys. Rev. Lett. \textbf{80}, 3843 (1998).

\bibitem{msr02} C. Panagopoulos, J. L. Tallon, B. D. Rainford, \ T.Xiang, J.
R. Cooper, and C. A. Scott, Phys. Rev. B \textbf{66}, 064501 (2002).

\bibitem{msr03} J. L. Tallon, J. W. Loram, C. Panagopoulos,
cond-mat/0211048, to appear in J. Low Temp. Phys.

\bibitem{Mon95} R. Monasson, Phys. Rev. Lett. \textbf{75}, 2875 (1995).

\bibitem{MP991} M. M\'{e}zard and G. Parisi, Phys. Rev. Lett. \textbf{82},
747 (1999). (1989).

\bibitem{SSW85} Y. Singh, J. P. Stoessel, and P. G. Wolynes, Phys. Rev.
Lett. \textbf{54}, 1059 (1985).

\bibitem{KTW89a} T. R. Kirkpatrick and P. G. Wolynes, Phys. Rev. A \textbf{35%
}, 3072 (1987).

\bibitem{KTW89b} T. R. Kirkpatrick and P. G. Wolynes, Phys. Rev. B \textbf{36%
}, 8552 (1987).

\bibitem{KTW89c} T. R. Kirkpatrick and D.Thirumalai, and P. G. Wolynes,
Phys. Rev. A \textbf{40}, 1045 (1989).

\bibitem{KTW89d} T. R. Kirkpatrick and D. Thirumalai, Phys. Rev. Lett. 
\textbf{58}, 2091 (1987).

\bibitem{TOE}R. B. Laughlin and D. Pines, Proc. of the Natl. Acad. of Sciences {\bf 97}  27 (2000).

\bibitem{RY79} T. V. Ramakrishnan and M. Yussouff, Phys. Rev. B \textbf{194}%
, 2775 (1979); M. Youssouff, Phys. Rev. B \textbf{23}, 5871 (1983).

\bibitem{Si} Q. Si and J. L. Smith, Phys. Rev. Lett. \textbf{77}, 3391
(1996).

\bibitem{CK90} R. Chitra \ and G. Kotliar, Phys. Rev. B \textbf{63}, 115110
(2001).

\bibitem{cluster2} M. H. Hettler, A. N. Tahvildar-Zadeh, M. Jarrell, T.
Pruschke, H. R. Krishnamurthy, Phys. Rev. B \textbf{58}, R7475 (1998).

\bibitem{cluster3} A. I. Lichtenstein and M. I. Katsnelson, Phys. Rev. B 
\textbf{63}, R9283 (2000).

\bibitem{cluster} G. Biroli and G. Kotliar, Phys. Rev. B \textbf{65}, 155112
(2002).

\bibitem{Lopatin} A. V. Lopatin and L. B. Ioffe, Phys. Rev. B {\bf 66}, 174202 (2002) 

\bibitem{KB70} G. Baym and L. P. Kadanoff, Phys. Rev \textbf{124}, 287
(1961); G. Baym, Phys. Rev \textbf{127}, 1391 (1962).

\bibitem{SW00} J. Schmalian and P. G. Wolynes, Phys. Rev. Lett. \textbf{85},
836 (2000).

\bibitem{BCK96} J.-P. Bouchaud, L. Cugliandolo, J. Kurchan, M. M\'{e}zard,
Physica A, \textbf{226}, 243 (1996).

\bibitem{Geissler} P. L. Geissler and D. R.Reichman, cond-mat/0304254

\bibitem{SWW00} H. Westfahl Jr., J. Schmalian, and P. G. Wolynes, Phys. Rev.
B \textbf{64}, 174203 (2001).

\bibitem{WSW02} H. Westfahl Jr., J. Schmalian, and P. G. Wolynes, preprint

\bibitem{CK93} L. F. Cugliandolo, J. Kurchan, Phys. Rev. Lett. \textbf{71},
173 (1993).

\bibitem{mc} W. G\"{o}tze, in \emph{Liquids, Freezing and Glass Transition},
ed. J.-P. Hansen, D. Levesque and J. Zinn-Justin (North-Holland, Amsterdam,
1991), p. 287.

\bibitem{Hansen} J. P. Hansen and I. R. McDonald, \emph{Theory of simple
liquids}, Academic Press, London, N.Y., San Francisco, 1976.

\bibitem{dmft1} D. Vollhardt, in Correlated Electron Systems, edited by V.J.
Emery (World Scientific, Singapore, 1993).

\bibitem{dmft2} A. Georges, G. Kotliar, W. Krauth, and M. J. Rozenberg, Rev.
Mod. Phys. 68, 13-125 (1996).

\bibitem{ED75} R. T. Deam and S. F. Edwards, Phil. Trans. R. Soc. \textbf{%
280A}, 317 (1976).

\bibitem{Paul} P. M. Goldbart, H. E. Castillo, and \ A. Zippelius, Adv.
Phys. \textbf{45}, 393 (1996).

\bibitem{KT89} T. Kirkpatrick and D. Thirumalai, J. Phys. A Math.-Gen. 
\textbf{22}, L49 (1989).

\bibitem{Kivelson} D. Kivelson, S. A. Kivelson, X. L. Zhao, Z. Nussinov, and
G. Tarjus, Physica A {\bf 219}, 27 (1995).


\bibitem{Stillinger83} F. H. Stillinger, J. Chem. Phys. \textbf{78}, 4654
(1983).

\bibitem{Deem94} M. W. Deem and D. Chandler, Phys. Rev. E \textbf{49}, 4268
(1994).

\bibitem{Brazovskii} S. A. Brazovskii, Zh. Exp. Teor. Fiz. \textbf{68}, 175
(1975) [Sov. Phys. JEPT \textbf{41}, 85 (2975)].

\bibitem{Leibler} L. Leibler, Macromolecules \textbf{13}, 1602 (1980).

\bibitem{Fredrickson} G. H. Fredrickson and E. Helfand, J. Chem. Phys. 
\textbf{87}, 697 (1987).

\bibitem{EK93} V. J. Emery and S. A. Kivelson, Physica \textbf{209C}, 597
(1993).

\bibitem{WWSW02} S. Wu, H. Westfahl Jr., J. Schmalian, P. G. Wolynes, Chem.
Phys. Lett. \textbf{359}, 1 (2002).

\bibitem{Hohenberg} P. C. Hohenberg and J. B. Swift, Phys. Rev. E \textbf{52}%
, 1828 (1995).

\bibitem{dAT78} J. R. L. de Alameida and D. J. Thouless, J. Phys. A \textbf{%
11}, 983 (1978).

\bibitem{ptcomment} $\varepsilon $ as defined in this paper differs from the
one of Ref.[31] by a factor $2$. 
\end{thebibliography}
\end{document}